\documentclass[12pt,preprint]{aastex}

\usepackage{lscape}

\shorttitle{Arp 220 Nucleus}
\shortauthors{D.L. Clements et al.,}
\slugcomment{DRAFT: 2002 Mar 28}

\begin{document}


\title{Chandra Observations of Arp 220: The Nuclear Source}


\author{D.L. Clements}
\affil{
Physics Dept., Imperial College, Prince Consort Road, London SW7 2BW, UK\\
Dept. Physics and Astronomy, Cardiff University, PO Box 913, Cardiff,
CF24 3YB, UK}
\email{d.clements@ic.ac.uk}

\author{J.C. McDowell}
\affil{Harvard-Smithsonian Centre for Astrophysics, 60 Garden Street,
Cambridge, MA 02138, USA}

\author{S. Shaked}
\affil{University of Arizona, Dept. Astronomy, 933 North Cherry Avenue,
Tucson, AZ 85721-0065, USA}

\author{A.C. Baker}
\affil{Dept. Physics and Astronomy, Cardiff University, PO Box 913, Cardiff,
CF24 3YB, UK}

\author{K. Borne}
\affil{Raytheon Information Technology and Sciences Services, NASA Goddard
Space Flight Centre, Greenbelt, MD 20771, USA}

\author{L. Colina}
\affil{lnstituto de Estructura de la Materia (CSIC), Serrano 121, 28006 Madrid, Spain}

\author{S. Lamb}
\affil{Departments of Physics and Astronomy, Centre for Theoretical
Astrophysics, Loomis Laboratory of Physics, 1110 W. Green Street,
University of Illinois at Urbana-Champaign, Urbana, IL 61801, USA}

\author{C. Mundell}
\affil{ARI, Liverpool John Moores University, Twelve Quays House,
Egerton Warf, Birkenhead, Wirral, Cheshire, CH41 1LD, UK}



\begin{abstract}
We present the first results from 60ks of observations of Arp 220 using the
ACIS-S instrument on Chandra. We report the detection of several sources
near the galaxy's nucleus, including a point source with a hard spectrum
that is coincident with the western radio nucleus B. This point source is mildly
absorbed ($N_H\sim 3\times 10^{22}\mbox{cm$^{-2}$}$) and has an estimated
luminosity of $4\times 10^{40}$ erg/s. In addition, a fainter source
may coincide with the eastern nucleus A. Extended hard X-ray emission
in the vicinity raises the total estimated nuclear 2-10 keV X-ray luminosity to
$1.2\times 10^{41}$ erg/s, but we cannot rule out a hidden AGN behind
columns exceeding $5\times 10^{24}\mbox{cm$^{-2}$}.$ We also detect
a peak of soft X-ray emission to the west of the nucleus, and a hard point
source 2.5 kpc from the nucleus with a luminosity of $6\times10^{39}$ erg/s.
\end{abstract}




\section{Introduction}
Ultraluminous Infrared Galaxies (ULIRGs) have quasar-like bolometric
luminosities ($>10^{12} L_{\odot}$) dominated by the far-infrared
(8--1000$\mu$m) part of the spectrum (Sanders \& Mirabel, 1996).
Almost all ULIRGs are interacting or merging galaxies
(Clements et al. 1996), possibly linking them to the
transformation of disk galaxies into ellipticals (eg. Wright et al, 1990;
Baker \& Clements, 1997). The
prodigious luminosity of ULIRGs is thought to be powered by a massive
starburst, a dust--buried AGN or some combination of the two. Despite a
decade of work we still have not been able to decide between these
paradigms. Various scenarios have also been suggested linking the
evolution of quasars with ULIRGs (eg. Sanders et al., 1988).
These suggest that part of the luminosity we see from some
ULIRGs originates in a dust--obscured AGN which later destroys or expels
the enshrouding material. Meanwhile, studies of the X-ray
background (Mushotzky et al, 2000) suggest that
dust--enshrouded AGN make a substantial contribution to its hard
component.  Such objects may also be linked (Trentham \& Blain, 2001;
Almaini et al., 1999) to the
recently discovered Cosmic Infrared Background (Puget et al. 1996;
Fixsen et al., 1998) and the objects
that contribute to it (Puget et al. 1999; Sanders 2000
and references therein).  As the most obscured
objects in the local universe, and as strong candidates for making the
CIB, ULIRGs are ideal local laboratories for studying many of these
issues. 

Arp 220 is the nearest ULIRG, having an 8-1000$\mu$m luminosity of
$\sim 1.2 \times 10^{12}L_{\odot}$ and a redshift of $z=0.018$. As such
it is an ideal target for ULIRG studies. The consensus since ISO is
that Arp 220 is powered by a massive burst of star formation rather than
an AGN (Sturm et al 1996), but the possibility of a heavily obscured AGN powering
the bulk of its emission remains (Haas et al 2001). The evolutionary
scenario linking ULIRGs to AGN also allows the possibility that a
weak, but growing, AGN may lie at the centre of Arp 220. While this may
not be energetically significant at the present time, it may grow to
prominence at later stages in the object's evolution. The plausibility
of such a scenario has been investigated by Taniguchi et al. (1999),
who show that it is quite possible for a massive black hole ($\sim 10^6
M_{\odot}$) to grow to $\sim 10^8M_{\odot}$ during the course of a
galaxy merger, and thus to be capable of powering a quasar. 

Signs of AGN activity can be sought with X-ray observations. The current
data for Arp 220 includes soft X-ray images from ROSAT (Heckman et al.
1996). These show extended X-ray emission associated with the H$\alpha$
nebula (Arribas, Colina \& Clements 2001), which are thought to be
produced by a superwind. However the overall soft X-ray luminosity is
small relative to the far-IR luminosity when compared to other
starbursts, and might allow room for some AGN contribution (Iwasawa,
1999). At higher energies, where an AGN would be more prominent, data is
available from HEAO-1 (Rieke, 1988), CGRO (Dermer et al., 1997), ASCA
(Iwasawa 1999), and BeppoSAX (Iwasawa et al. 2001). These rule out the
possibility of an unobscured energetically significant AGN in Arp 220. 
The possibility remains, however, of a Compton thick AGN, with an
obscuring column in excess of 10$^{25}$cm$^{-2}$, or of a weaker lower
luminosity AGN that will grow into a quasar.  We have thus undertaken
Chandra X-ray observations of Arp 220 aimed at detecting a weak or
obscured AGN in its nucleus, and to study the extended superwind
emission in detail. This paper presents the first results from our study
of the nuclear regions. Our results on the superwind can be found
elsewhere (McDowell et al. 2002, Paper II). We assume a distance of 76
Mpc (Kim \& Sanders 1998) to Arp 220 throughout this paper.

\section{Observations}

Chandra observed Arp 220 with the ACIS-S instrument for 58 ks on  2000 Jun
24.   The ACIS-S instrument was chosen for its good soft response to
allow us to study the low energy X-ray emission of the superwind, as
well as the harder emission expected from any nuclear source. We chose
to use the back-illuminated CCD S3, for maximum soft response and to
avoid any charge transfer difficulties arising in the front-illuminated
chips. Arp 220 is sufficiently faint that no pile-up issues were expected
or found in the data. The data were reduced by the standard Chandra
pipeline through level 1 (calibrated event list) and further analysed
using the CIAO package\footnote{http://cxc.harvard.edu/ciao} version 2.1 and 2.2.
The data were taken with the chip at a temperature of -120C and were
gain-corrected using acisD2000-01-29gainN0003.fits from the July 2001
recalibration. The observation was relatively unaffected by background
flares and only a small amount of exposure was removed, leaving  an
effective exposure time of 55756s. Astrometry was corrected using a
revised geometry file (telD1999-07-23geomN0004.fits) which is believed
to provide positions across the full ACIS field accurate to about 1
arcsecond.
The standard screening (good time intervals and grade filtering for grades
0,2,4,5,6) was applied to generate a cleaned event file.
The X-rays from Arp 220 extend over 20 kpc (Paper II), but
emission above 2 keV is restricted to the central few kpc.

\section{The Nuclear Source}

\subsection{Imaging the Nuclear Source}

\begin{figure}
\epsscale{1.0}
\caption{\small X-ray color image of the nuclear region of Arp 220, reconstructed
by adaptive smoothing. Red: 0.2-1 keV; Green: 1-2 keV; Blue: 2-10 keV.
White bar 2 arcseconds long is provided for scale. The bright
source in the blue area is X-1, and the one in the yellow area is X-3.
X-2 is the blue source to lower right, while X-4, just to lower left of
X-1, is too faint to be visible in this picture.\label{fig1}}
See file 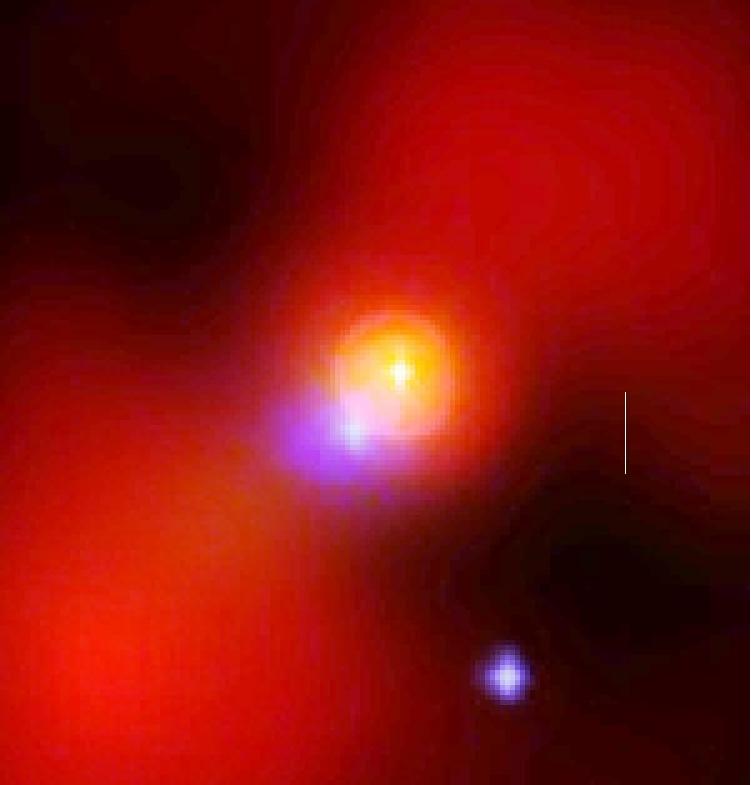 for colour figure.
\end{figure}

Figure \ref{fig1} is a true X-ray color image of the Arp 220 nuclear region.
It was smoothed in separate bands of 0.2-1 (red), 1-2 (green) and 2-10 keV
(blue) using the CIAO adaptive smoothing routine {\it csmooth}.
The image shows that the nuclear region of Arp 220 is clearly
distinguished from the rest of the object by being the site of much
harder emission.

The centroid of the soft emission is displaced 1.5 arcseconds to the
northwest of the hard emission. The hard emission coincides with a dust
lane in the galaxy (Joy et al. 1986), and indeed the soft emission is
suppressed there. However, the absence of hard emission away from the
nucleus shows that the spectral change is due to a different type of
source, and not merely an absorption effect.

\begin{figure}
\epsscale{1.0}
\plotone{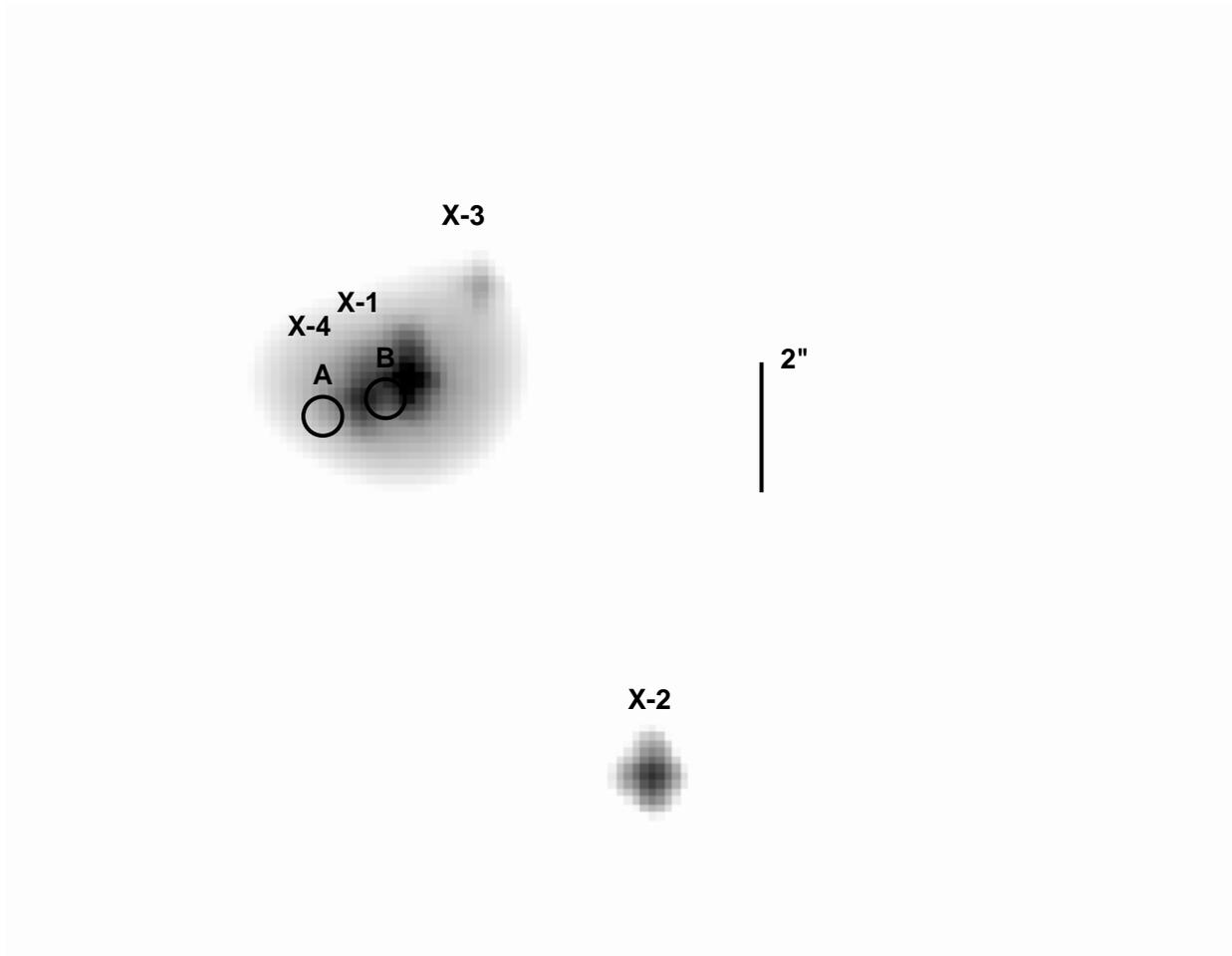}
\caption{\small Hard X-ray (2-10 keV) image. The positions of
sources X-1 to X-4 are indicated; the circles A and B are the
positions of radio nuclei. 2 arcsecond bar provided for scale. No attempt has been made to
register the two frames; the Chandra astrometric accuracy is about 1 arcsecond. \label{fig2}}
\end{figure}

Figure \ref{fig2} shows an image of the hard emission
($>4$ keV) coming from the nuclear regions of Arp 220, together with
circles indicating the area within 1'' of the well-studied dual radio
and IR nuclei (see eg. Scoville et al. 1998). The positional match between the
radio/IR nuclei and the hard emission is $\sim$1'', within the
expected pointing accuracy of Chandra. (The surprising lack of detections of USNO
stars in the field limits our ability to improve the astrometric accuracy,
but three galaxies are found within one arcsecond of their published positions.)

Previous observations of Arp 220 have shown the presence of
hard emission, using, for example, Beppo/SAX (Iwasawa et al.
2001). However, it is only with the angular resolution of Chandra that
we have been able to localise some of this emission to the region of
the nuclei.

The mean off-axis angle of the nucleus during the observation was only
38 arcseconds, so the point spread function (calculated using the
standard CIAO tools) is very close to the on-axis value.  To estimate
source fluxes, we generated monochromatic point spread functions for the
 midpoints of the three energy bands using the CIAO tool mkpsf, and
subtracted a point source from the raw images at the location of the
hard peak, with the maximum amplitude that did not create a dip in the
local diffuse flux when the data was then smoothed. This procedure was
repeated at the location of remaining flux peaks.

The central region is consistent (within the 1" absolute
astrometric accuracy of the Chandra data) with
a pair of hard point sources at the positions of radio nuclei A and B
(Scoville et al 1998), with the western nucleus dominant, although
the best fit separation for two point sources is only 0.7" rather than
the 1.1" of the radio positions and the decomposition of this small region
into multiple point sources and diffuse emission is not unique.

We designate the sources in order of total flux. The hard band image
shows emission concentrated around a nuclear source, X-1, with extended
emission around it. 
Point source subtraction suggests the presence of a much weaker second
hard nucleus, X-4, together with a diffuse component which we denote as
the X-1 halo; its centroid is 0.5 arcseconds east of X-1.
X-2, further out from the nucleus, is a hard source detected out to 5
keV; only 33 net counts are seen. Its luminosity of $6\times 10^{39}\mbox{erg s$^{-1}$}$
 puts it in the interesting category of non-nuclear ULX (ultra-luminous
X-ray) sources. 
X-3, the soft peak, is not reliably separated from the extended soft
circumnuclear emission, and coincides with the peak of the H$\alpha$
emission (Arribas, Colina \& Clements 2001; McDowell et al. 2002).

The 2--10keV luminosity we derive for the nuclear source X-1 and
surrounding hard emission (assuming the spectral model derived
in the next section) is 6.9$\times
10^{40}$ ergs s$^{-1}$, measured from a 3'' aperture centred on the
nucleus. Because of the poorly constrained absorption, here and below we quote
observed luminosities rather than absorption-corrected ones unless
explicitly stating otherwise.
The hard luminosity found here compares to the 2--10keV luminosity found by Iwasawa et
al. (2001), in a 3 arcminute aperture, of 11$\times 10^{40}$ ergs s$^{-1}$,
assuming a similar distance to Arp 220.  The background in our
observation makes us  insensitive to hard emission on 3 arcminute
scales, and we cannot rule out an extended contribution of this
magnitude. However, the 3 sigma upper limit to any remaining hard (2-8
keV) flux within 1 arcminute of the nucleus is $2.6\times10^{40}\mbox{erg s$^{-1}$}$
and we speculate that emission from the nearby southern group may be
contributing to the Iwasawa et al result. Deeper observations with
XMM/Newton would resolve this issue.

\subsection{Spectral Analysis}

\begin{figure}
\epsscale{0.7}
\plotone{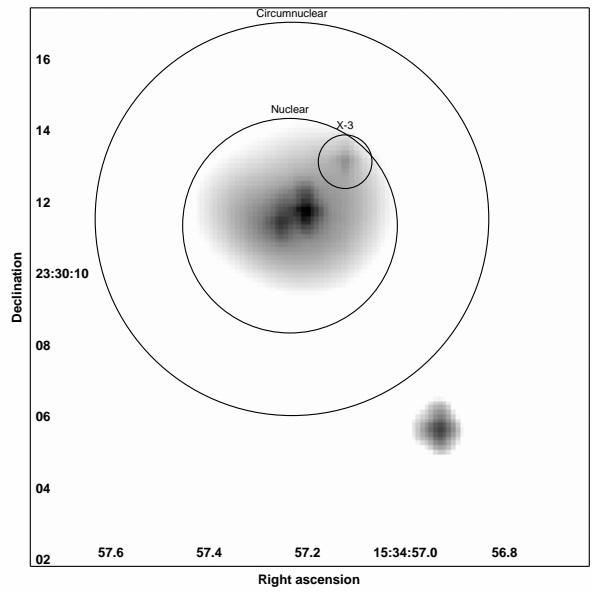}
\caption{\small Extraction regions for spectral analysis,
superimposed on 2-10 keV image. Coordinates are J2000.
\label{fig3r}}
\end{figure}

We considered two extraction regions: a 5.5" radius `circumnuclear'
region, including X-3 but with the X-1/X-4 region omitted; a
`nuclear' region with X-1/X-4 and the hard halo, taking photons within
a 3'' radius circle, but excluding photons within 0.75" of
the X-3 location to minimize contamination by the soft source.  The
extraction regions are indicated  in figure \ref{fig3r}.  For background
we extracted counts from two 50" radius circular regions 3' away on
either side of the galaxy and lying on the same node of the chip.

Responses were generated using the CIAO tools {\it mkwarf}
and {\it mkrmf}; a PI (gain-corrected) spectrum was extracted and grouped by a
factor of 10 (0.15 keV bins). Both XSPEC and Sherpa were used
to fit models to the data, and gave similar results.

\begin{figure}
\epsscale{0.7}
\plotone{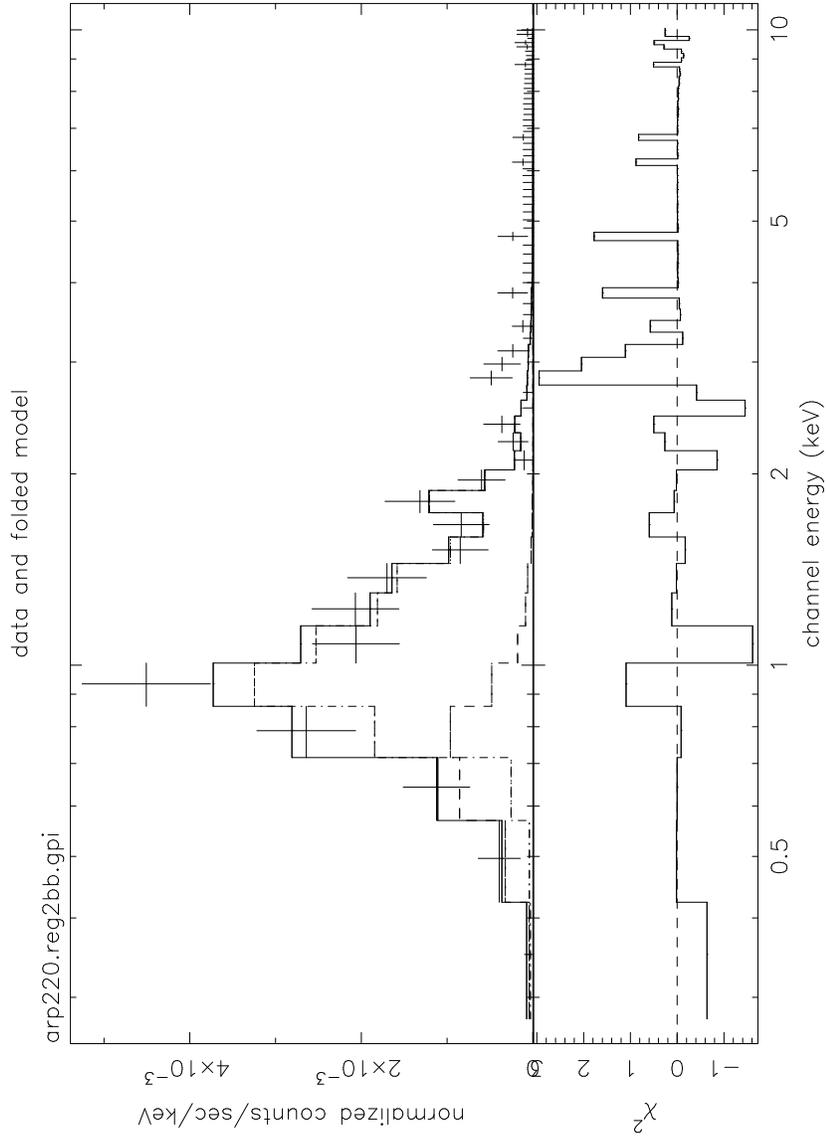}
\caption{\small Spectrum of circumnuclear region, including
soft X-ray peak but excluding region of hard emission. Data
were grouped in 150 eV bins and fit with the sum of two
Raymond-Smith thermal models. Lower panel shows contribution
to $\chi^2$ in each bin.
\label{fig3}}
\end{figure}

The circumnuclear emission is fit with a Raymond-Smith or Mekal model with 
a two-temperature plasma of 0.14 and 0.9 keV and absorption of
$0.3-1 \times10^{22}\mbox{cm$^{-2}$}$. It is probably composed of a mixture of diffuse
emission and point sources. There is no sign of hard
emission in this spectrum, confirming the impression from the images that
hard emission is restricted to the central kiloparsec of Arp 220.
The data are shown in figure \ref{fig3}. 

\begin{figure}
\epsscale{0.7}
\plotone{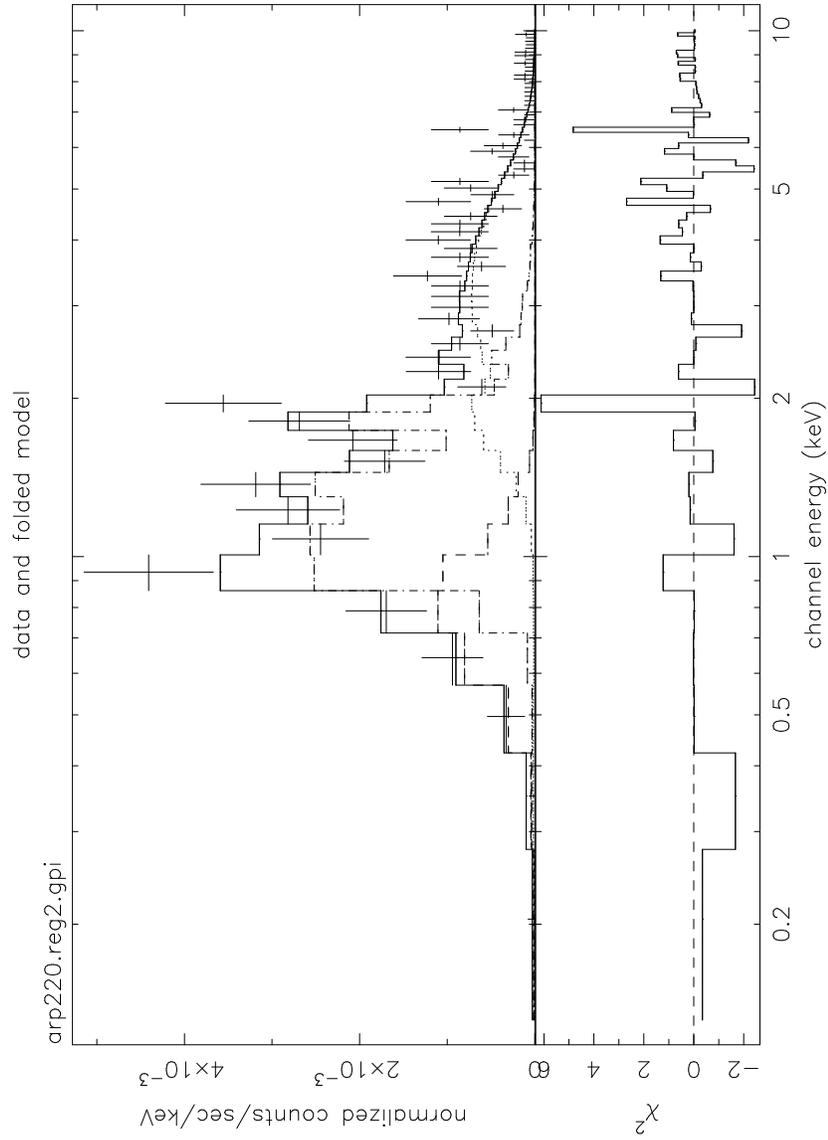}
\caption{\small Spectrum of nuclear region, showing hard power law
and multicomponent fit. Extraction is from 3 arcsecond radius circle
around X-1, but excludes a 1 arcsecond region around X-3.
\label{fig4}}
\end{figure}

Fitting the region of hard nuclear emission we find a spectrum similar
to the soft circumnuclear emission but with  an additional power law
component.
A representative fit has a power law component
of photon index 1.4$\pm$1 behind $N_H=3\pm2\times10^{22}\mbox{cm$^{-2}$}$. 
The absorbed 0.3-10 keV
luminosity of the power law component is $6.9\times10^{40}\mbox{erg s$^{-1}$}$,
and most of the counts are above 2 keV. The thermal components,
assumed to be superimposed circumnuclear emission, have twenty percent
of the flux of the power law, so the hard component dominates the
energetics of the nuclear emission. The overall fit, with 9 free
parameters, has a reduced chi-squared of 0.84 for 59 d.o.f and
the C-statistic is 62.4;
see figure \ref{fig4}.


If instead of modelling the soft emission directly, we use the
off-nuclear circumnuclear emission as a background, we find a fit to the
remaining emission consistent with a simple power law of index
1$\pm$0.5, a lower NH around $0.4\times 10^{22}\mbox{cm$^{-2}$}$, and a similar absorbed
luminosity of $5.6\times10^{40}\mbox{erg s$^{-1}$}$. However, the remaining emission
is also consistent with the slope 1.4 power law plus a modest amount of
additional thermal emission at 0.9 keV.

The thermal models are not good at fitting the 2.0 keV (rest) Si XIV
line; an extra contribution at 1.95 keV (rest)  with L=$9\times10^{38}\mbox{erg s$^{-1}$}$
reduces the $\chi^2$, suggesting a possible slight gain error.

There is a high bin near 6.5 keV which could be a weak Fe K line. To
evaluate this possibility we inspected the raw gain-corrected pulse
height data without binning. There are a mere 10 photons with
gain-corrected energies in the 0.4 keV wide bin between 6.3 and 6.7 keV,
compared to 7 photons in the 0.8 keV on either side between 5.9-6.3 and
6.7-7.1, a 2 sigma `detection'. The distribution of the counts is
consistent with the instrumental width at that energy, and their spatial
distribution suggests a distribution centered towards the eastern source
X-4 (although a tighter spatial selection only improves the detection
significance to 2.6 sigma). If real, this line would have a luminosity
of $4\times 10^{39}\mbox{erg s$^{-1}$}$, an equivalent width of 700 eV and a rest
energy of $6.62\pm0.05$ keV; these may be taken as approximate upper
limits to any narrow Fe K line.


The absorption-corrected luminosity of the nuclear emission, including the
hard halo, is $1.3 (1-2.6)\times10^{41} \mbox{erg s$^{-1}$}$.
Based on the PSF subtraction calculations, we assign 
30 percent of this to X-1, giving $\sim 4\times 10^{40}\mbox{erg s$^{-1}$}$
which is in the range of both ultra-low luminosity
AGN and ultra-luminous binaries (Fabbiano 1998, King et al. 2001).
See Table 1 for derived source positions and luminosities.

Could there be a further hard source at this location, behind a much
larger column? Using a canonical photon index of 1.7, we set limits of
L(0.2-10 keV) = $1.4\times 10^{42}\mbox{erg s$^{-1}$}$ for a column of $10^{24}
\mbox{cm}^{-2}$, and $2.5\times 10^{44}$ for a column of $5\times
10^{24}\mbox{cm$^{-2}$}$. However, for a column of $10^{25}\mbox{cm$^{-2}$}$, which is entirely
plausible in the center of Arp 220, no useful limit can be set by the
ACIS data as even the most luminous quasars would have their X-ray flux
absorbed; one must turn to the harder X-ray limits from the BeppoSAX PDS
of $2\times10^{42}\mbox{erg s$^{-1}$}$ in 13-50 keV (Iwasawa et al. 2001) to eliminate
this possibility. In general, because of the low sensitivity of Chandra
above 6 keV, the limits we set  as a function of absorbing column are
weaker than those in Figure 5 of Iwasawa et al. for the range they study
(log $N_H$=24.3 and above).


\section{Discussion}

\subsection{Spatial Distribution of Nuclear X-ray Emission}

The X-ray emission in Arp 220 is clearly divided into two parts --- the
compact, hard nuclear emission, and the diffuse, softer extended
emission. This diffuse emission clearly extends right
into the nucleus since the thermal component of the nuclear emission
appears very similar to the thermal emission found in the off-nuclear
spectrum. The hard emission is not extended beyond a
region $\ll$1kpc in size (3'' at 76 Mpc, our adopted distance for
Arp 220).  This makes an interesting contrast with other starburst and
interacting/merging galaxies already studied by Chandra. For example,
in the Antennae (Fabbiano, Zezas \& Murray 2001) hard emission comes from numerous point
sources extended over a 10kpc region, and in NGC 3256 (Ward et al. 2000), we see
similar clumps spread over 3.5kpc. If such sources existed in Arp 220,
we would expect to detect all those above $\sim$5$\times 10^{39}$ergs
s$^{-1}$ cm$^{-2}$ in luminosity. We only see one such source, X-2, which
lies 7'' away from the nucleus. The hard X-ray emission from Arp 220
thus appears to be significantly more concentrated in the nucleus than
that of other interacting or merging galaxies observed by
Chandra. Indeed, the spatial distribution of hard emission in Arp 220
would appear to be more similar to that of Mrk273 (Xia et al. 2002), a ULIRG
containing an AGN at its core, than NGC3256 or the Antennae, neither
of which seem to have a significant AGN contribution.

\subsection{Spectrum of the Arp 220 Nuclear Region}

The X-ray output from the nuclear regions of Arp 220 is energetically
dominated by an extended hard kpc-scale component, with a significant
point source contribution ($<200$pc), that has a power law spectrum with
$\Gamma$=1.4$\pm$1.0, with an unabsorbed
2-10keV luminosity of 4$\times 10^{40}$ ergs/s.
The central question is the origin of this
radiation -- does it arise from young supernovae, X-ray binaries, or
the result of accretion onto a more massive body, possibly a
weak AGN. We consider each of these possibilities in turn:
\begin{itemize}
\item {\bf Young Supernovae} Young supernovae can be strong X-ray emitters
especially if in a dense circumstellar or interstellar environment
(Schlegel, 1995). Such objects are also strong radio emitters. 
Colina et al (2001) detected a luminous radio supernova in
the nuclear starburst of the Seyfert galaxy NGC 7469, a marginally
ultraluminous infrared galaxy. 12
candidates for such 'radio supernovae' have been detected by VLBI/VLBA
observations in the nuclei of Arp 220 (Smith et al. 1998), so we know
such objects are present.
The X-ray properties of such objects are
highly heterogeneous. SN1986J, for example, had an integrated
0.1-2.4keV luminosity of 2--3$\times$10$^{40}$ergs s$^{-1}$ cm$^{-2}$,
while SN1993J has a luminosity in the same band of
3$\times$10$^{38}$ergs s$^{-1}$ cm$^{-2}$. The temporal behaviour of
these objects are also different, ranging from relatively constant
emission over several years for SN1978K to decay times of a few weeks
or months for SN1993J (Schlegel, 1995). A number of such objects
are capable of powering the emission we see in Arp 220, though some
long term variability should be detectable, especially since no new
radio supernovae appear to have occurred for some years (Lonsdale et al. 2000).
Spectrally the emission from young supernovae are best fitted by
thermal bremsstrahlung spectra. Attempts to fit the nuclear spectrum of Arp 220 
with a  thermal bremsstrahlung model give an implausibly high
temperature of 10.5 keV with a 1.7$\times 10^{22}$cm$^{-2}$
absorbing column ($\chi^2$=52.4 with 77 degrees of freedom);
supernovae typically have T$\sim$0.5keV. Fixing
the emission to this temperature gives a fit with a poorer $\chi^2$
than other models, with $\chi^2$ = 83 with 78 degrees of freedom.

\item {\bf X-Ray Binaries and ULX sources} Typical galactic X-ray binaries have luminosities
up to $\sim$10$^{37}$ergs s$^{-1}$ cm$^{-2}$ (White, Nagase \& Parmar 1995),
so several
thousand would be required to produce the luminosity seen in Arp 220's
nucleus, which is implausible. 
However, galaxies such as the Antennae and NGC3256, with
considerably more active star formation than our own galaxy, have been
found to contain ultra-luminous X-ray sources (ULX) with luminosity comparable
to Arp 220. A survey of 11 nearby spirals with Chandra finds 15 ULX sources
(Kilgard et al, in preparation). The ULX sources may be
accreting objects with mass $\geq
500M_{\odot}$, as seen in M82 (Griffiths et al. 2000, Kaaret et al. 2001); or, the
objects may be high-mass X-ray binaries with compact object masses in the
stellar range, but with beamed emission (King et al. 2001).
The M82 source appears to have a
10keV thermal bremsstrahlung spectrum, which can fit the data well
(see above), so the presence of several ULX sources in the core of Arp 220
is feasible. 

\item{\bf Inverse Compton Emission}
Moran et al. (1999) suggested that the hard X-ray emission in NGC3256
is produced through inverse Compton scattering of the far-IR photons
by relativistic electrons produced by supernova remnants in the
starburst. Conditions suitable for inverse Compton emission might also
exist in the nuclei of Arp220, and give rise to the hard emission seen
here. One test of inverse Compton emission is that the radio and X-ray
power law indicies should match, and this does indeed prove to be the
case. However, as noted by Iwasawa et al. (2001), the inverse Compton
process in Arp220 must be quite inefficient or the hard X-ray luminosity
would be much greater.

\item{\bf Active Galactic Nuclei}
The restriction of much of the hard X-ray emission to just the central
regions of Arp 220 is naturally explained if the source is an AGN
rather than an agglomeration of XRBs. However, whilst the nuclear
regions of Arp 220 have a high luminosity for an individual XRB, they
have a low luminosity for an AGN, being about 2 orders of magnitude
fainter than typical Seyfert 1's (George et al. 1998). However,
luminosities this low are not without precedent for suspected AGNs. Ho
et al. (2001) have conducted a survey of nearby galaxies with Chandra,
examining the X-ray emission of objects that might contain low
luminosity AGNs. They find weak AGNs in a substantial fraction of
their sample, with X-ray luminosities (2-10keV) as low as
$10^{39}$ergs s$^{-1}$ cm$^{-2}$. They also find a correlation between
X-ray and H$\alpha$ luminosity which Arp 220 fits on when only the
nuclear H$\alpha$ luminosity is considered (Armus, Heckman \& Miley
1990). If there is a weak AGN such as this in Arp 220, it will
contribute only about 10$^8$---10$^9$L$_{\odot}$ to the bolometric
luminosity, ie. less than 1\%. However, the presence of even a weak
AGN in the dense environment of the nucleus of Arp 220 raises the
possibility that it will grow and significantly increase in luminosity
as the system continues to evolve.  There is also the possibility that
we are only seeing a small fraction of the emission from an AGN if the
source itself is Compton obscured. The tentative detection of an
emission line at 6.5keV, with equivalent width $\sim$1 keV, raises the
possibility that the power-law source seen is just a reflected
component from an otherwise obscured source. Harder X-ray observations
by Beppo/SAX (Iwasawa et al. 2001) set the best limits to date for any
obscured AGN. A moderately Compton thick AGN, with 2$\times 10^{24}$
cm$^{-2}$ column, according to these limits, could supply a few
percent of the bolometric luminosity, assuming standard
L$_{x}$/L$_{bol}$ ratios. Absorption by $>$10$^{25}$ cm$^{-2}$ is
needed before a substantial fraction of Arp 220's luminosity can come
from an AGN.  Such extinction is not impossible. Haas et al. (2001)
use the 7.7$\mu$m/850$\mu$m flux ratio to suggest very high
extinctions, while the LWS spectrum from Fischer et al. (1999),
indicating $\tau>1$ at 100$\mu$m, would imply an HI column of
2.7$\times 10^{25}$cm$^{-2}$.  Improved sensitivity spectra for the
nuclear regions of Arp 220 is clearly needed to confirm whether the
6.5keV line is real and if it is associated with reflected emission.
The presence of {\em extended} hard X-ray emission in the nuclear
regions of Arp220 cannot be easily explained by a simple AGN model.
Possible explainations would include a mixed nature for the nuclear
regions, with unresolved AGN emission combining with extended XRB
emission in the halo.  Such an arrangement would naturally follow from
Taniguchi et al.'s (1999) model of nuclear black hole growth by
accretion of XRBs during a merger.  Alternatively, an obscured AGN
might be associated with extended scattered hard emission. The
acquisition of deeper Chandra images to examine any differences in
X-ray properties between unresolved and halo hard X-ray emission would
be able to address these possibilities.

\end{itemize}

\section{Conclusions: The Nature of Arp 220}

We have shown that there is a source of hard, power-law-like, X-ray
emission in the nuclear regions of Arp 220. This source is extended
EW, consistent with the emission coming from both the radio/IR
nuclei. The central concentration of hard X-ray emission in Arp 220 is
in contrast to other interacting galaxies, where hard emission comes
from clumps distributed across much larger physical distances. This
difference may be associated with the merger in Arp 220 being older
than in the Antennae or NGC3256, and that compact objects have sunk to
its core (Tremaine, Ostriker \& Spitzer, 1975) but may also be
associated with Arp 220's greater luminosity. The origin of the hard
emission is unclear.  Its spectrum is unlikely to be produced by young
supernovae, but inverse Compton emission, albeit of very low
efficiency, accretion onto ultra-luminous X-ray binaries or onto an
AGN are all possible. If there is an AGN contribution, it has too low
a luminosity for it to play a significant role in the energetics of
the object, but the presence of even a weak AGN in Arp 220 would
support a connection between ULIRGs and quasars (Sanders et al. 1988).
If it is not an AGN, we cannot rule out the presence of a true AGN
behind a Compton screen of column $\sim10^{25}\mbox{cm$^{-2}$}$.  If
XRBs are responsible for the emission then a large number of
conventional XRBs are needed, or a smaller number of ultraluminous ULX
sources.  In either case, the concentration of these objects in the
very centre of Arp 220 might indicate that these objects will later
merge together to form a supermassive black hole and AGN (Taniguchi et
al. 1999). If there is a weak AGN already in the nuclear regions, then
this process may already have begun.

\acknowledgments

We acknowledge use of the ADS, NED, and CIAO.

Partial support for this work was provided by the National Aeronautics
and Space Administration through Chandra Award Number GO1-1166 issued by
the Chandra X-Ray Observatory Center, which is operated by the
Smithsonian Astrophysical Observatory for and on behalf of NASA under
contract NAS8-39073. S. Lamb acknowledges support from the DOE,
through LLNL.  DLC and ACB were supported by PPARC, CM by the Royal
Society. We would also like to acknowledge the contributions of Bernie
Peek and Charlie Baker to this work.

\clearpage


\begin{landscape}
\begin{table*}
\caption{Flux of Arp 220 X-ray Components}
{
\small

\smallskip

\begin{tabular}{llllllll}
\\
Object                       & Position               & Extent$^a$  & Net   & F$_{14}^b$ & $L_{40}^c$ & $N_H^d$ & $L_{40}(0)^e$ \\
                             & (J2000)                & ($\arcsec$) & counts&(obs)    &  (erg s$^{-1}$)    & ($10^{22} \mbox{cm}^{-2}$) & (erg/s) \\
\hline\noalign{\smallskip}
Arp 220 Circumnuclear        & 15:34:57.14 +23:30:13.0  &  7"      & 250& 1.6 & 1.1 & 0.5 (0.3-1)& 6\\ 
Arp 220 X-1 hard halo        & 15:34:57.28 +23:30:11.4  &  3"      & 113& 5.9 & 4.1 & 3 (1-5)    & 8? \\ 
Arp 220 X-1 (nucleus)        & 15:34:57.21 +23:30:11.7  &  Unres.  &  66& 3.0 & 2.1 & 3 (1-5)    & 4? \\ 
Arp 220 X-2                  & 15:34:56.94 +23:30:05.5  &  Unres.  &  33& 1.0$^f$ & 0.6 & 0.3 (0.2-0.6)&0.7 \\  
Arp 220 X-3                  & 15:34:57.14 +23:30:13.1  &  2"      &  31& 0.2$^g$ & 0.1 & 0.5 (0.3-1) & 0.6\\ 
Arp 220 X-4                  & 15:34:57.25 +23:30:11.5  &  Unres.  &  19& 1.0$^h$ & 0.7 & 3 (1-5)     & 1.5?\\ 
\end{tabular}

\parbox{5.5in}{
$^a$ Estimated extent of source in arcseconds.
$^b$ Observed flux in the 0.3-10.0 keV band in units of $10^{-14} \mbox{erg cm$^{-2}$ s$^{-1}$}$.\\
$^c$ Luminosity corresponding to observed flux (uncorrected for absorption)\\
in units of $10^{40}$ erg/s.\\
$^d$ Hydrogen column in units of $10^{22}$ cm$^{-2}$: best value and range.\\
$^e$ Unabsorbed luminosity in units of $10^{40}$ erg/s corresponding to correction
for absorption using best value from previous column. Question marks indicate
that the absorption correction is highly uncertain.\\
$^f$ Luminosity estimated using a power law model fit with the photon index
fixed at 1.0.\\
$^g$ Luminosity found assuming the circumnuclear two-temperature model
fit for the circumnuclear region.\\
$^h$  Luminosity estimated by assuming the same spectral model used for X-1.\\
}

}
\end{table*}
\end{landscape}


\clearpage

\end{document}